\documentclass[amsmath,twocolumn,showpacs,amssymb,aps,nofootinbib]{revtex4-1}
\usepackage{graphicx}
\usepackage{subcaption}
\usepackage{pgfplots}
\usepackage{dcolumn}
\usepackage{bm}
\usepackage{slashed}
\usepackage{amsmath}
\usepackage{amssymb}
\usepackage{tikz}
\usepackage{bbold}
\usepackage{mathtools}
\usepackage{romannum}
\usepackage{natbib}
\usepackage{braket}
\usepackage{appendix}
\newcommand{\be}{\begin{equation}}
	\newcommand{\ee}{\end{equation}}
\newcommand{\ba}{\begin{eqnarray}}
	\newcommand{\ea}{\end{eqnarray}}

\newcommand{\red}{\textcolor{black}}
\usepackage[pagebackref=false, colorlinks=true]{hyperref}
\definecolor{redish}{rgb}{0.7,0.2,0.0}  
\definecolor{bluish}{rgb}{0.2,0.5,0.8}
\hypersetup{linkcolor=red,          
	citecolor=teal,        
	filecolor=magenta,      
	urlcolor=blue}          

\begin{document}
\pagenumbering{arabic}
\title{Time evolution of spread complexity and statistics of work done in quantum quenches \\}
\author{Kuntal Pal}
\email{kuntal@iitk.ac.in}
\author{Kunal Pal}
\email{kunalpal@iitk.ac.in}
\author{Ankit Gill}
\email{ankitgill20@iitk.ac.in}
\author{Tapobrata Sarkar}
\email{tapo@iitk.ac.in}
\affiliation{
Department of Physics, Indian Institute of Technology Kanpur-208016, India}
\date{\today}
\begin{abstract}
We relate the probability distribution of the work done on a statistical system under a sudden quench to the Lanczos coefficients corresponding to 
evolution under the post-quench Hamiltonian. Using the general relation between the moments and the cumulants 
of the probability distribution, we show that the Lanczos coefficients can be identified with physical quantities associated with the 
distribution, e.g., the average work done on the system, its variance, as well as the  higher order cumulants. In a sense this gives 
an interpretation of the Lanczos coefficients in terms of experimentally measurable quantities. Consequently, our approach provides a
way towards understanding spread complexity, a quantity that measures 
the spread of an initial state with time in the Krylov basis
generated by the post quench Hamiltonian, from a thermodynamical 
perspective.
We illustrate these relations with two examples. The first one involves quench done on a harmonic chain with periodic boundary 
conditions and with nearest neighbour interactions. As a second example, we consider mass quench in
a free bosonic field theory in $d$ spatial dimensions in the limit of large system size. In both cases, we find out the time evolution
of the spread complexity after the quench, and relate the Lanczos coefficients with the cumulants of the 
distribution of the work done on the system.

\end{abstract}
\maketitle
\section{Introduction}

Physical quantities that are used to probe non-equilibrium scenarios such as a sudden  quench of a quantum mechanical
system  \cite{Polkovnikov:2010yn} are broadly 
categorised into two different sets. The first includes different correlation functions \cite{Igloi, Calabrese1}, entanglement entropy \cite{Calabrese2}, 
out of time order correlators, as well as the more recently introduced notions of complexity (studied in the context of quantum quenches  in e.g. \cite{Alaves, 
Camargo, Caputa2, spread1}). Time evolution of these quantities after a quench generally show some characteristic behaviour at early as well as late times, 
and these can also be used to detect criticality in many body quantum systems that exhibit quantum  phase transitions.
The second set is related to thermodynamics. Indeed, it has been established that the problem of quantum quenches can be viewed as a thermodynamic 
transformation \cite{Barankov:2008qq, Polkovnikov}. This fact offers a new way of looking at the physics after a quantum quench in terms of 
quantities that are commonly used to characterise standard thermodynamic processes, such as the heat and entropy generated,
as well as  the work done on the system \cite{Silva}. 
 
In this work, our main focus will be a new measure of the complexity of constructing a target wave function starting from a reference one, 
namely the spread complexity (SC), introduced in \cite{Balasubramanian:2022tpr}. The complexity of a state is broadly 
a measure of the minimum number of basis gates one requires to construct that state, starting from a given 
reference state. There are various notions of this measure, such as the Nielsen complexity (NC) \cite{Nielsen1,Nielsen2,Nielsen3,Jefferson:2017sdb, Khan:2018rzm, Hackl:2018ptj, Bhattacharyya:2018bbv}, 
Fubini-Study complexity (FSC) \cite{Chapman:2017rqy}, Bi-invariant complexity \cite{Yang:2018cgx}, complexity from covariance matrix
\cite{Chapman:2018hou}, complexity from information geometry \cite{DiGiulio:2021oal}, path integral approach to circuit optimization \cite{Bhattacharyya:2018wym}, 
possible extensions to conformal field theories \cite{Erdmenger:2020sup,  Basteiro:2021ene} etc. (see the recent review \cite{Chapman:2021jbh} and references therein).   
These studies have gained a lot of attention in the recent literature due to usefulness of various notions complexity
as probes of   quantum phase transitions 
\cite{Liu:2019aji, Xiong:2019xoh, Jaiswal:2020snm, Jaiswal:2021tnt, Sood:2021cuz, Sood:2022lfx, Gautam:2022gci, Huang:2021xjl,  
Roca-Jerat:2023mjs, Jayarama:2022xta, Pal:2022ptv, Pal:2021jtm}, 
quantum quenches \cite{Camargo, Bhattacharyya:2018wym, Ali:2018aon, Ali:2018fcz, 
Chandran:2022vrw, Pal:2022rqq, Choudhury:2022xip, Adhikari:2021pvv}, as well  as indicators of quantum chaotic evolution \cite{Ali:2019zcj, Balasubramanian:2019wgd, Qu:2022zwq}.

In a similar vein, the SC of a state under the evolution by a unitary operator measures the \text{spreading} of 
the wavefunction on the Hilbert space \cite{Balasubramanian:2022tpr}. Intuitively the more it spreads over the corresponding Hilbert space, 
the more difficult it is to construct that state. The central idea of finding the SC of a state lies in the
construction of the Krylov basis by the Hermitian operator that generates the flow, which is done by the
 well-known Lanczos algorithm of constructing a tri-diagonal form of a given matrix \cite{Viswanath, Parker:2018yvk, Lanczos:1950zz}. 
 This process takes the auto-correlation function of the final state and the initial state 
as an input, and gives two sets of coefficients, known as the Lanczos coefficients (LC) as outputs. Using 
  the discretised version of the Schrödinger equation on the Krylov basis, it is then possible to obtain the 
  SC as the minimum value of the associated cost for the Krylov basis, as was proven in 
  \cite{Balasubramanian:2022tpr}. There is a lot of recent attention on various aspects of the SC and 
  its corresponding operator version (Krylov complexity), see for examples works related to quantum 
  phase transition \cite{Caputa2, Caputa:2022yju, spread1}, operator scrambling 
  \cite{Barbon:2019wsy, Bhattacharjee:2022vlt}, conformal field theory \cite{Dymarsky:2019elm,
  Dymarsky:2021bjq, Kundu:2023hbk}, open systems \cite{Bhattacharya:2022gbz, Bhattacharjee:2022lzy, Bhattacharya:2023zqt}, 
as a tool of probing delocalization properties in nonchaotic quantum systems \cite{Kim:2021okd}, and other related 
contexts \cite{Yates:2021asz, Caputa:2021ori, Patramanis:2021lkx, Trigueros:2021rwj, Rabinovici:2021qqt,  
Rabinovici:2022beu, Bhattacharjee:2022qjw, Chattopadhyay:2023fob, Kundu:2023hbk, Bhattacharjee:2023dik, 
Bhattacharjee:2022ave, Takahashi:2023nkt, Camargo:2022rnt, Avdoshkin:2022xuw, Erdmenger:2023shk, Alishahiha:2022anw, Muck:2022xfc, Adhikari:2022whf, Adhikari:2022oxr}.

In this paper our goal is to relate two of these quantities from the above mentioned different sets, 
thereby offering a new way of interpreting the evolution 
 of a system after a quantum quench. In particular, the two apparently distinct quantities that we consider in this paper are (a) the 
 LC used to study the SC, and (b) various cumulants of the probability distribution of the 
 work done on the system by suddenly changing its parameters. That these two sets of numbers are related to each other can, in some sense, be 
``guessed'' by noting that the characteristic function (CF) associated with the distribution of the work done is related to the 
 complex conjugate of the auto-correlation function --  a quantity whose moments contain all the information
 about the LC. In this paper we make this relation precise and quantify this with two examples.   
 
The rest of the paper is organised as follows. In section \ref{worklanczos}, we first briefly review the two quantities mentioned above, 
and then obtain a relation between the LC and the cumulants of the probability distribution by using Faa di Bruno's formula. 
We show that these quantities are related to each other via the Bell polynomials. In section \ref{singlequench}, we apply this 
formalism to the time evolution of the SC after a single sudden 
quench of the parameters of a harmonic chain with periodic boundary conditions,
and provide a physical interpretation of the LC when a critical quench is considered. Section \ref{bosonic}
elaborates on our second example -- the mass quench of a noninteracting bosonic model in $d$-spatial dimensions in the limit
of infinite linear size of the system. Section \ref{conclusions} discusses the main outcomes of our analysis.

\section{Statistics of work and the Lanczos coefficients in quantum quenches}\label{worklanczos}

In this section we first briefly review the  formalism of the statistics of work done under a sudden  quantum
quench and the Lanczos algorithm of obtaining the  LC from the auto-correlation function
of a time-evolved state.

\subsection{Statistics of the work done under a quantum quench}

It is well known that  a quantum quench can be viewed as a thermodynamic transformation 
\cite{Barankov:2008qq, Polkovnikov}. Thus, apart from the 
quantities related to time evolution of quantum correlation functions and  quantum information
theoretic quantities (such as the entanglement entropy or complexity), a fundamental way to characterise
quantum quench is to consider the statistics of the work done on a quantum system when its parameters are changed
suddenly \cite{Silva}. The work done on the system through a given quench protocol is defined as the 
difference between the internal energies before and after the quench.
That we need a probability distribution function to quantify the work done on the system
can be understood from the fact that due to the sudden change of the parameters of the system Hamiltonian,
even if we consider different realisations of the same quench protocol, the measurement of the  work done
will yield different results i.e., it will show fluctuations.

To mathematically quantify the distribution of the work done, we  consider the simplest single
quench protocol, where the parameters of a quantum system, collectively denoted by $\lambda$, are 
changed suddenly to a new set of values $\tilde{\lambda}$ at an instant of time $t=t_0$ (we will usually take $t_0=0$).
We denote the energy eigenstate before and after the quench as $\big| n (\lambda) \rangle$ and 
$\big| \tilde{n} (\tilde{\lambda}) \rangle$, which correspond to energies $E_{n}(\lambda)$ and $\tilde{E}_{n}(\tilde{\lambda})$,
respectively. The Hamiltonian before and after the quench are denoted by $H_0$ and $\tilde{H}$.

Now, if energy measurements before and after the quench give the 
results $E_{m}$ and $\tilde{E}_n$ respectively, then the probability distribution of the work done $W=\tilde{E}_n-E_m$
is given by \cite{Kurchan, Talkner1}
\begin{equation}\label{statistics}
P(W)=\sum_{m,n}\big|\langle  m\big|\tilde{n}\rangle\big|^2 \delta \big(W- (\tilde{E}_n-E_m)\big)~.
\end{equation}

For the quench protocols considered in this paper, we shall take the state before the quench to be the 
ground state of the Hamiltonian. Hence the above distribution reduces to 
\begin{equation}\label{statistics2}
P(W)=\sum_{n}\big|\langle 0\big|\tilde{n}\rangle\big|^2 \delta \big(W- (\tilde{E}_n-E_0)\big)~.
\end{equation}
Once we know the distribution of the work, it is useful to first consider its CF defined as the
Fourier transform 
\begin{equation}\label{CF}
G(t)=\langle e^{-i W t}\rangle=\int e^{-i W t} P(W) dW~.
\end{equation}
The importance of this quantity in the context of quantum quenches, as established in \cite{Talkner1},  is that
the CF is actually the correlation function
\begin{equation}\label{CF2}
\begin{split}
G(t)&=\langle 0\big|e^{i H_{0} (\lambda)t}e^{-i \tilde{H} (\tilde{\lambda})t}\big|0\rangle~
=e^{i E_{0} (\lambda)t}\langle 0\big|e^{-i \tilde{H} (\tilde{\lambda})t}\big|0\rangle~\\&=e^{i E_{0} (\lambda)t} \langle \psi_0\big|\Psi(t)\rangle~,
\end{split}
\end{equation}
where $ \big|\psi_0\rangle = \big|0\rangle $ is the state before the quench at $t=0$. Now it is easy to see that 
this quantity is just the conjugate of the Loschmidt echo studied  extensively in the context of the quantum quenches
and quantum chaos \cite{Peres, Jalabert, Goussev, Gorin}. Furthermore, we can see that it is related to the complex
conjugate of the auto-correlation function, apart from a trivial phase factor. This is the first indication of a
connection between the LC and the CF, which we elaborate on below. 

For future references, at this point it is useful define the  cumulants ($\beta_n, n\geq 1$) of the distribution as 
expansions of the logarithm of the CF\footnote{To be consistent with the notion of the quench and spread complexity
literature, we have defined the cumulant expansion with factors of $i$.}
\begin{equation}\label{cumulantex}
\ln \big(G(t)\big) = \sum_{n=1}^{\infty} \frac{(-it)^n}{n!}\beta_n~, ~~~
\beta_n=(-i)^{-n} \frac{d^n \ln G(t)}{dt^n}\Big|_{t=0}~.
\end{equation}
Furthermore, we can also write down the expansion of the CF in terms of the moments ($M_n$) of the distribution 
\begin{equation}\label{momentex}
G(t) = \sum_{n=0}^{\infty}\frac{t^n}{n!}M_n~.
\end{equation}
Note that since $G(t=0)=1$, here we have $M_0=1$ by definition. Furthermore, the way we have defined the coefficients, $M_n$s are actually related to average of $W^n$ with powers of 
$i$, i.e. $M_n=(-i)^n\int W^n P(W)dW$. Using this along with the expression in Eq. \eqref{statistics2} for a quench
from the ground state, we obtain the following meaningful expression for the moments in terms of products of the energy
difference between energy eigenstates before and after the quench and their overlap
\begin{equation}\label{genmoments}
M_n=(-i)^n \sum_{j}\big|\big<0\big|\tilde{j}\big>\big|^2 (\tilde{E}_j -E_0 )^n~.
\end{equation}
Here, $\big|\tilde{j}\big>$ represents an eigenstate of post-quench Hamiltonian with energy $\tilde{E}_j$.
Therefore, we see that the moments depend on the overlap between the initial state (here, the ground state 
of the pre-quench Hamiltonian) and the eigenstates of the post-quench Hamiltonian, as well as the spacing between 
energy levels of the post-quench Hamiltonian with that of the ground state energy of the initial Hamiltonian
$H_0$. 
In appendix \ref{homoment} we illustrate an example of this formula, where we obtain the moments 
 for the case of a quench in a single harmonic oscillator.

It is well known that the cumulants of a probability distribution
 are related to its central moments. For example, the first cumulant ($\beta_1$) is the mean of the distribution
 (here $ \langle W\rangle $), the second cumulant is its 
 variance $\sigma^2$ (i.e. the second central moment; here $ \langle W^2\rangle- \langle W\rangle ^2 $ ), and the third cumulant  is
 equal to the third central moment, etc. All the other higher cumulants are actually polynomial functions of the central moments
with integer coefficients.

The behaviour of these quantities -- the probability distribution of the work done (PDWD) $P(W)$, CF 
$G(t)$, and the moments -- have been studied after sudden quenches in systems which show quantum phase transitions 
e.g., in \cite{Silva} and \cite{Paraan}, by taking the Ising chain and the Dicke model as prototypical examples.
In these references it was established that the moments of the  PDWD show diverging behaviour 
when the system is quenched through the critical points of the systems. See also the 
recent works \cite{Fei1, Fei2}, where the authors have studied the work distribution
of a quench across  a quantum phase transition, and found universal scaling relations in such cases.

 In this paper, we  relate 
the moments of the distributions with the LC and study the properties of these moments (and the 
cumulants) of the distribution when there is a zero mode present in the dispersion relation, as well as in the limit
of infinite system size.

\subsection{Lanczos coefficients and the Krylov basis construction}
We now briefly review the Lanczos algorithm of constructing the Krylov basis and the definition of the 
SC of an arbitrary initial state under Hamiltonian evolution. 

The central idea behind the construction of the Krylov basis is to write the Hamiltonian in the tri-diagonal basis in the Lanczos algorithm. 
In this construction, a new basis is defined from the old one as follows : 
\begin{equation}\label{Krylov-basis}
\big| {K_{n+1}} \rangle=\frac{1}{b_{n+1}}\Big[\big(H-a_{n}\big)\big|{K_n}\rangle-b_{n}\big|{K_{n-1}}\rangle\Big]~.
\end{equation}
We take $\big|{K_{0}}\rangle=\big|{\psi(0)}\rangle$ i.e.  the algorithm starts from the reference state. 
The computation of the coefficients $a_{n}, b_{n}$ (known as the LC) plays a key role in implementing the Lanczos algorithm. 
It is important to note that information about the LC is also encoded in the so called `return-amplitude,' which is
 defined as the overlap 
between the state at any particular value of the circuit parameter $t$ and the initial state, i.e.,
\begin{equation}
\mathcal{S}(t)=\langle \psi(t)|\psi(0)\rangle~.
\end{equation}
Once we have constructed the Krylov 
basis for the Hamiltonian evolution, we can expand the desired state in this basis as 
\begin{equation}
\big|{\psi(t)}\rangle=\sum_{n}\phi_{n}(t)\big|{K_{n}}\rangle~.
\end{equation}
It can be shown that the expansion coefficients $\phi_{n}(t)$ satisfy the following  discrete Schrodinger equation
\begin{equation}\label{dse}
i\dot{\phi}_{n}(t)=a_{n}\phi_{n}(t)+b_{n}\phi_{n-1}(t)+b_{n+1}\phi_{n+1}(t)~,
\end{equation}  
where a dot denotes a derivative with respect to $t$. It was recently proved that the Krylov basis as defined above, 
minimises the cost function $\mathcal{C}_{B}(t)=\sum_{n}n|\langle {\psi(t)|B_{n}}\rangle|^2$,  which
measures the spreading of the state under the desired evolution   
\cite{Balasubramanian:2022tpr}. Here, $\big|B_{n}\rangle$ is the particular basis which we use to evaluate the spreading. 
We can write the above cost in the Krylov basis as 
\begin{equation}\label{spread_complexity}
\mathcal{C}(t)=\sum_{n}n|\phi_{n}(t)|^2~.
\end{equation}
This  is the definition of the SC.

Next we briefly describe how the LC can be calculated from the return amplitude, and subsequently, how
the  $\phi_{n}(t)$ are obtained by solving Eq. (\ref{dse}). 
For calculating the LC from the return amplitude, we first need to find the even and odd moments 
from the expansion
\begin{equation}\label{moments}
\mathcal{S}(t)=\sum_{n}^{\infty}M_{n}^{*}\frac{t^n}{n!}~.
\end{equation}
Here, $M_{n}^{*}$s are the expansion coefficients of the return amplitude. After knowing the  moments, we can find the 
full sets of $a_{n}$s and $b_{n}$s using the standard recursion methods available in the literature for dynamics under Hermitian evolution \cite{Viswanath, Lanczos:1950zz}, that was recently extended for the case of open systems in \cite{Bhattacharjee:2022lzy}, which we briefly recall below. 

To construct the full set of orthonormal Krylov basis on the Hilbert space, we  start from the given state $\big|{\psi(0)}\rangle$, i.e., 
this is the first Krylov state $\big|{K_{0}}\rangle=\big|{\psi(0)}\rangle$. Then the recursion relation of Eq. (\ref{Krylov-basis}) implies that the 
next basis is  $\big|{K_{1}}\rangle=\frac{1}{b_{1}}\big[(H-a_{0})\big|{K_{0}}\rangle\big]$. Here we have used the fact that $b_{0}=0$. The condition that 
this state $\big|{K_{1}}\rangle$ is orthogonal to the previous state $\big|{K_{0}}\rangle$ fixes the unknown coefficient $a_{0}$ to be equal to 
$\langle {K_{0}|H|K_{0}}\rangle$. The other coefficient $b_{1}$ ensures the normalisation of this state. We continue this recursive process to 
construct the full set of basis and the general coefficients are  given as 
\begin{equation}
a_{n}=\langle {K_{n}|H|K_{n}}\rangle~,
\end{equation}
while $b_{n}$s fix the normalization at each step. However in practice, where the above process does not terminate after first few steps, it is 
more useful to implement the Lanczos algorithm by means of two sets of auxiliary matrices $L_{k}^{(n)}$ and $M_{k}^{(n)}$ constructed from the 
moments $M_{n}^*$s of the return amplitude defined in Eq. (\ref{moments}). The recursion relations then can be written down in terms of those  
$L_{k}^{(n)}$s and $M_{k}^{(n)}$s and finally the LC are obtained as $b_{n}=\sqrt{M_{n}^{(n)}}$ and $a_{n}=-L_{n}^{(n)}$ 
with the initial conditions properly chosen ($b_{0}=0$) \cite{Balasubramanian:2022tpr, Viswanath}. This is a  standard procedure, and we refer the reader to \cite{Viswanath} for details. 
Once we have the full set of LC, we have all the information that is required to find the coefficients $\phi_{n}$, by solving the discrete Schrodinger 
equation in Eq. (\ref{dse}),  and calculate the spread complexity as a function of time.

\subsection{Relation between the Lanczos coefficients and the cumulants}

From the discussions of the previous two subsections, it should be clear that we can relate the LC 
associated with the Lanczos algorithm (and the Krylov basis), with fundamental physical quantities characterising
a sudden quench as a thermodynamic transformation -- such as the average work, its variance and higher
order cumulants. To
establish such a relation, we need to relate the cumulants of the expansion (which have information of the 
average work, variance etc) with the moments  (from  which we can obtain the LC).
This can be achieved using Faa di Bruno's formula which generalizes the chain rule to the higher derivatives
(see e.g., \cite{Fraenkel}).
Here we briefly outline the derivation of this relation to use the notation consistent with the ones used above and the 
rest of this section. For details of the derivation, we refer to,   e.g., \cite{Lukacs, Lloyd}

First we define the partial Bell polynomials as the coefficients in the expansion of the following generating function
of two variables \cite{Comtet}
\begin{equation}\label{partialBell}
\begin{split}
\Phi (t,u)&=\exp \bigg(u\sum_{m=1}^{\infty} g_m \frac{t^m}{m!}\bigg)~\\
&=~1+\sum_{n=1}^{\infty}\sum_{k=1}^{n} B_{n,k} \big(g_1, \cdots g_{n-k+1}\big)u^{k}\frac{t^n}{n!}~.
\end{split}
\end{equation}
The partial Bell polynomial $B_{n,k}$ is a homogeneous polynomial of degree $k$ and weight $n$ in the expansion 
coefficients $g_m$. 
Evaluating this at $u=1$ we get the definition of the complete Bell polynomials $Y_{g_1,\cdots g_n}$ as
\begin{equation}\label{completeBell}
\Phi (t,1)=\exp \bigg(\sum_{m=1}^{\infty} g_m \frac{t^m}{m!}\bigg)~
=~1+\sum_{n=1}^{\infty} Y_{n} \big(g_1, \cdots g_{n}\big)\frac{t^n}{n!}~,
\end{equation}
so that the complete polynomials are related to  the partial polynomials  through
\begin{equation}
Y_{n} \big(g_1, \cdots g_{n}\big) = \sum_{k=1}^{n} B_{n,k} \big(g_1, \cdots g_{n-k+1}\big)~.
\end{equation}
Explicit expressions for both the partial and complete Bell polynomials are known which are written as a sum over 
all the partitions of $n$ into $k$ non-negative parts, and all the partitions of $n$ into arbitrarily many
non-negative parts respectively \cite{Lloyd}. 

Now using the Faa di Bruno's formula we can obtain 
\begin{equation}
\begin{split}
&\ln \bigg(\sum_{n=0}^{\infty} M_n \frac{t^n}{n!}\bigg)= \ln M_0 + \\
&\sum_{n=1}^{\infty}\sum_{k=1}^{n} (-1)^{k-1} (k-1)! M_0 ^{-k} B_{n,k} \big(M_1, \cdots M_{n-k+1}\big)\frac{t^n}{n!}~,
\end{split}
\end{equation}
where we have assumed $M_0>0$. In fact, since we are considering the expansion of the CF around the start of the quench at $t=0$, 
here $M_0=1$ and the first term in the above expansion vanishes. We can now compare the right hand side with the cumulant 
expansion in Eq. \eqref{cumulantex} to obtain the desired relation between the cumulant of the PDWD 
($\beta_n$) and the moments $M_n$ which carry the information of the LC as
\begin{equation}\label{relation}
\beta_n =  \sum_{k=1}^{n} \frac{(-1)^{k-1}}{(-i)^n} (k-1)!  B_{n,k} \big(M_1, \cdots M_{n-k+1}\big)~.
\end{equation}
This formula will be used in the next section to relate the average and variance of the work done on the system through
a quench and the LC (such as $a_0$ and $b_1$ etc) in a simple manner.

 It is also useful to obtain the inverse of the above relation. This 
can straightforwardly obtained using the definition of the complete Bell polynomials given in Eq. \eqref{completeBell} to be 
\begin{equation}\label{invrelation}
M_n = (-i)^n Y_n (\beta_1, \cdots \beta_n)~.
\end{equation}
The well known tabulated expressions for the Bell polynomials can now be used  to  explicitly relate these two types of 
expansion coefficients.  

Once we know these moments in terms of the cumulants, they can be used, following the 
	procedure outlined in the previous subsection, to obtain the LC. For example, we  have  the following relations
	between $a_0$ and the average work, and 
	$b_1$ and the variance of the work done
	\begin{equation}\label{a0b1}
	a_0=iM_1=\big<W\big> ~, ~~\text{and}~~ b_1^2=-M_2+M_1^2=\big<W^2\big>-\big<W\big>^2~.
	\end{equation}
	Similarly, it is possible to write the coefficient $a_1$ in terms of $\big<W^3\big>$, variance and the average of the work done 
	done, and is given by 
	\begin{equation}\label{a_1}
	a_1=\frac{\big<W^3\big>-\big<W\big>^3}{\big<W^2\big>-\big<W\big>^2}-2\big<W\big>~.
	\end{equation}
See Appendix \ref{a_1der} for a brief derivation of the last two equations.
	Other LC can similarly be written in terms of averages of $\big<W^n\big>$, however, their expressions become 
	increasingly cumbersome, and hence we avoid writing them here. 

Since we have established that all the LC can be related to the moments and cumulants of the PDWD, 
at this point, it is important to comment on the measurements of PDWDs in experimental setups. It is clear from its definition in Eq. \eqref{statistics}, that
to measure the work distribution, one needs to perform two projective energy measurements on the system, one before the quench and the other after it. 
It is well known that it is difficult to perform reliable projective measurements of energy in
many-body quantum systems. However for relatively simple systems, there exist
pioneering experiments where  the quantum  work statistics have been measured. Examples of such systems include, a spin-1/2 system 
undergoing closed nonadiabatic  evolution (which can be realised in NMR setups) \cite{Batalhao}, driven oscillator systems which can be used to describe the dynamics of trapped ions  
\cite{An}, or ultracold atoms \cite{Cerisola}. Since the relations between the LC and the work distribution we have obtained are very general (valid for any evolution
with any generic time-independent Hamiltonian), our results can be applied to these cases as well.
Therefore, in a sense, relations of the kind given in Eqs. \eqref{a0b1}, \eqref{a_1} can be thought of
as providing an interpretation of LC in terms of experimentally measurable quantities.

Before concluding this section, here we mention an important relation between a universal behaviour of the survival probability
at short times after quench  ($t<<\sigma^{-1}$) and $b_1$, where $\sigma$ is the variance of the PDWD
under the quench (see below for its definition). The expression
for the  survival probability, defined as the modulus 
square of the overlap between the initial state and the time-evolved state, is given by
\begin{equation}
\mathcal{P}(t)=\big|\langle \psi(0)|\psi(t)\rangle\big|^2=\Big|\sum_{n}\big|\big<0\big|\tilde{n}\big>\big|^2
e^{-i\tilde{E}_n t}\Big|^2~.
\end{equation} 
Now from Eq. \eqref{genmoments}, we first write down  the expressions for the 
average and the variance of the PDWD under quench in terms of the overlaps of pre and post-quench energy eigenstates:
\begin{align}
\big<W\big>=\sum_{j}\big|\big<0\big|\tilde{j}\big>\big|^2 (\tilde{E}_j -E_0 )~,\\
\sigma^2=\sum_{j}\big|\big<0\big|\tilde{j}\big>\big|^2 \Big(\tilde{E}_j -E_0 -\big<W\big>\Big)^2~.
\end{align}
Next, we expand the above expression for the survival probability, and for times $t<<\sigma^{-1}$  obtain \cite{Santos}
\begin{equation}
	\mathcal{P}(t) \approx 1-\sigma^2 t^2~ \approx 1-b_1^2 t^2~.
\end{equation}
Therefore, as is well known, for a very short time after a quench, the survival probability shows  universal quadratic 
decay in $t$ \cite{Santos}. Importantly, here we see that the rate of  decay  of the survival probability is actually determined
by $b_1$. Since the time behaviour shown by the survival probability  is universal, this observation shows 
the important role played by $b_1$ in the early time evolution of a quenched quantum many-body system.

\subsection{Lanczos coefficients in quench of a general system of length $L$ in $d$ dimensions}\label{ddimsys}

Before moving on to describe particular examples of the formalism discussed till now, in this subsection 
we consider a fairly general case of quench in a closed system placed in a 
$d$-dimensional box of length $L$ with periodic boundary conditions. We assume that a sudden  quench is
performed in such a system prepared in an initial state at zero temperature.
Work statistics and its CF for quenches done  in such scenarios have been studied previously in 
\cite{Gambassi, Sotiriadis, Palmai}.
 
In such a quench scenario, the expressions for the PDWD and its CF are the one given in Eqs. (\ref{statistics2})
and (\ref{CF2}) respectively.  Now consider performing a Wick rotation from time $t$ to the imaginary time
$-i\tau$  on the amplitude $\langle 0\big|e^{-i \tilde{H} t}\big|0\rangle$. The transformed amplitude
$\langle 0\big|e^{- \tilde{H} \tau}\big|0\rangle$ can be thought as the partition function of a  $(d+1)$
dimensional classical system with Hamiltonian $\tilde{H}$, defined on a strip of thickness $\tau$ \cite{Gambassi}.
The boundaries of the strip are described by the boundary states  $\big|0\rangle$. Using techniques used 
in the studies of the  critical Casimir effect, it is possible to  evaluate partition functions of such geometries,
so that continuing back to real time $\tau=it$, the expression for the CF can be conveniently written as the
sum of three terms
\begin{equation}\label{CF3}
\ln G(t) = -L^d \big[i f_b t + 2f_s + f_{C}(it)\big]~.
\end{equation}
Here, in the first term, which is linearly dependent on time, 
 $f_b=\frac{\tilde{E}_0-E_0}{L^d}$ is the difference between the ground state energies 
of the  Hamiltonians after and before the quench, per unit volume. Clearly, this term comes from a linearly time dependent
overall phase factor to the CF. On the other hand the second term, which is the surface free energy associated with 
each of the two boundaries of the strip geometry, is time-independent. This quantity is actually 
related to the fidelity between the ground states of the Hamiltonians before and after the quench through the 
relation $f_s=-L^d \ln \big|\langle \tilde{0} \big|0\rangle\big|$ \cite{Gambassi,Sotiriadis}. The third term in Eq. (\ref{CF3}) has non-trivial time dependence,
 and represents an effective interaction between two boundaries. 
 
Using the general formula for the CF written above, let us now calculate the cumulants, and hence relate the corresponding 
LC to the probability distribution. We have the expression for the cumulants as 
\begin{equation}\label{cumulants}
\begin{split}
\beta_n = \left\{\begin{array}{ll}-i L^d \Big[i f_b +\frac{d f_c (it)}{dt}\Big|_{t=0}\Big]~~~~\text{for}~~~n=1\\
         -i^n L^d \frac{d^n f_c (it)}{dt^n}\Big|_{t=0}  \quad \quad \quad~~ \text{for}~~~~n \neq 1.~\end{array}\right.
\end{split}
\end{equation}
Since in the second term, the contribution of the surface free energy of the boundary is time independent, it does not contribute 
to the expressions for the moments. Thus the ground state fidelity or its higher order derivatives, which are widely used to 
study  quantum quenches and quantum phase transitions are insensitive to the behavior of the  LC in a 
quench problem.  Furthermore, the first term has only an additive  contribution in $\beta_1$, i.e the first LC, $a_0$.
In most cases, we can omit the first term since one usually measures the work done $W$ starting from $\delta E_{gs}=\tilde{E}_0-E_0$
\cite{Sotiriadis}. 
In fact, most of the literature on the Lanczos algorithm which use the auto-correlation function (the
complex conjugate of the CF defined  in Eq. \eqref{CF2}), neglects the constant phase factor in the definition
of the CF in Eq. \eqref{CF2}.  In the example of quench in a harmonic chain considered in next section we will also
neglect this phase factor.

Thus the cumulants, moments of the PDWD, and hence the LC of the quench problem under consideration
are determined by the time dependence of the third term in the expression for the CF, namely, $f_c (it)$.
If we  now assume  that time derivative of this function is zero for $n=k+1$, then $\beta_{k+1}=0$,
and as we can see from the explicit expressions for the complete Bell polynomials $Y_n (\beta_1, \cdots \beta_n)$, 
all the higher order moments are determined by the first $\beta_k$ cumulants.
In  section \ref{bosonic}, we will discuss the limit $L \to \infty$ of the quench problem considered
in this section, and  explicit expressions for the time dependence of the function $f_C$  will be obtained 
to find out the cumulants and moments of the work distribution.

\section{Lanczos coefficients, spread complexity and statistics of work in quench of a harmonic chain}
\label{singlequench}

In this section we first discuss the time evolution of the SC after a single sudden quench of the
parameters of a harmonic chain with periodic boundary conditions. 
Then we study  the relation between the PDWD
under the quench and the corresponding LC, with particular emphasis given on the case of the critical quench,
i.e., when the final (or the initial) frequency of the oscillators of the harmonic chain vanishes.
  
The protocol for the quench we consider is the following. At $t=0$, we change the initial Hamiltonian (denoted as $H_0$) to 
a new one $H_1$, which has different values of the frequency and interaction strengths than $H_0$. Subsequent 
evolution the system is governed by the new Hamiltonian $H_1$.
We calculate the SC by taking the state at $t=0$ as the initial state i.e., the first state of the Krylov basis, and the target state 
as the time-evolved state after the  quench. For our purposes, in this section we assume that the reference state, i.e.,  
the state before the quench is the ground state of the  initial Hamiltonian $H_0$.

We denote the Hamiltonian before the  quench (i.e., for $t<0$) as $H_0$, which is given by
\begin{equation*}
\begin{split}
 H_0=\frac{1}{2}\bigg[\sum_{j=1}^{N}\Big(p_{0j}^2+\lambda_{0}^2x_{0j}^2\Big)
 +\sum_{j=1}^{N}k_0\Big(x_{0j}-x_{0(j+1)}\Big)^2\bigg]\\
 =\frac{1}{2}\bigg[\sum_{j=1}^{N}p_{0j}^2+X_0^T\cdot \mathcal{K}_0 \cdot X_0 \bigg]~.
\end{split}
\end{equation*}
Here $\lambda_{0}$ is the frequency of each oscillator before the quench, and $k_0$ is the nearest neighbour 
interaction strength. Furthermore,  $X_0=(x_{01},x_{02},...x_{0N})^T$ denotes the column matrix for the collective position of each oscillator, and $\mathcal{K}_0$ is 
a real symmetric matrix whose eigenvalues are denoted as $\omega_{0j}$. It is assumed that
periodic boundary conditions are imposed on the chain. We diagonalise this Hamiltonian  by performing
an orthogonal transformation $U_0$  which changes the coordinates to $Y_0=(y_{01},y_{02},...y_{0N})^T=U_0X_0$.
Denoting the transformed momenta as $P_{0k}$, the diagonal form for the Hamiltonian is therefore given by
\begin{equation}\label{Hamil0}
\begin{split}
&H_0(y_{0k},P_{0k})=\frac{1}{2}\sum_{k=1}^{N}\Big[P_{0k}^2+\omega_{0k}^2 y_{0k}^2 \Big]~\\
&=\sum_{k=1}^{N} \omega_{0k}\Big[a_{0k}^\dagger a_{0k}+\frac{1}{2}\Big]~
= \sum_{k=1}^{N} H_{0k}(y_{0k},P_{0k},t),
\end{split}
\end{equation}
where, $\omega_{0k}^2=\lambda_{0}^2+2k_{0}\big[1-\cos \big(\frac{2 \pi k}{N}\big)\big]$ is the $k$ th normal mode frequency, 
and in the second equality we have introduced the usual creation and annihilation operators for the individual modes.
  We assume that by the quench, only the frequency of all the oscillators $\lambda_{0}$ are changed simultaneously
to a new value $\lambda_{1}$, while keeping the interaction strength fixed.
After the quench, the  Hamiltonian of the harmonic chain can similarly be diagonalised in terms of a new set of
creation and annihilation operators as
\begin{equation}\label{Hamil1}
\begin{split}
&H_1(y_{1k},P_{1k})=\frac{1}{2}\sum_{k=1}^{N}\Big[P_{1k}^2+\omega_{1k}^2 y_{1k}^2 \Big]~\\
&=\sum_{k=1}^{N} \omega_{1k}\Big[a_{1k}^\dagger a_{1k}+\frac{1}{2}\Big]~=\sum_{k=1}^{N}H_{1k}(y_{1k},P_{1k})~,
\end{split}
\end{equation}
where the expressions for the normal mode frequencies are now given by 
\begin{equation}
\omega_{1k}^2=\lambda_{1}^2+2k_{1}\Big[1-\cos \Big(\frac{2 \pi k}{N}\Big)\Big]~,
\end{equation}
where $k_1$ is the nearest neighbour interaction strength after the quench, and here we assume that 
it is fixed, i.e. $k_1=k_0$  for  the quench protocol we consider.

The time-evolved state at an arbitrary time $t$  after  the quench is given by
\begin{equation}
\big| \Psi(t)\rangle=  e^{- i H_1 t} \big| \Psi_0\rangle =  e^{- i H_1 t} \big|K_0\rangle~.
\end{equation}
Here $\big| \Psi_0\rangle$ is the state before the first quench, and as we have mentioned before, this is the first state of 
the Krylov basis.  Here we take the ground state of the initial Hamiltonian $H_0$ to be the initial state 
$\big| \Psi_0\rangle=\big| 0\rangle=\prod_{k=1}^{N} \big| 0\rangle_k$. 

Since we have separated the Hamiltonian into individual modes  (see Eqs. (\ref{Hamil0}) and (\ref{Hamil1})), we can write the time-evolved state
after the quench as a product of  each individual time-evolved modes  
\begin{equation}
\big| \Psi(t)\rangle_k=  e^{- i H_{1k} t} \big| 0\rangle_k~,
\end{equation}
so that the SC of the state $\big| \Psi(t)\rangle$ is the sum of individual SCs of $N$ such modes. Below we illustrate 
the calculation of SC of an individual time-evolved mode $\big| \Psi(t)\rangle_k$.

\subsection{Auto-correlation function}
Since the annihilation and creation operators before the quench i.e., $(a_{0k}, a_{0k}^\dagger)$ respectively, are related to the operators after
the quench $(a_{1k}, a_{1k}^\dagger )$ through Bogoliubov transformations, the post-quench  Hamiltonian
 $H_{1}$ can be expressed in the following way in terms of the pre-quench operators  as  \cite{spread1,Ali:2018fcz}
\begin{equation}
H_1=2 \sum_{k=1}^{N} \omega_{1k} \big[\mathcal{U}_{1k}\mathcal{V}_{1k}\mathbf{K}_k^+ +\big(\mathcal{U}_{1k}^2
+\mathcal{V}_{1k}^2\big) \mathbf{K}_{k}^{0}+\mathcal{U}_{1k}\mathcal{V}_{1k}\mathbf{K}_k^-\big]~,
\end{equation}
where the Bogoliubov coefficients $\mathcal{U}_{jk}$ and $\mathcal{V}_{jk}$ are given by
\begin{equation}
\mathcal{U}_{1k}=\frac{\omega_{1k}+\omega_{0k}}{2 \sqrt{\omega_{1k} \omega_{0k}}}~,
~~\mathcal{V}_{1k}=\frac{\omega_{1k}-\omega_{0k}}{2 \sqrt{\omega_{1k} \omega_{0k}}}~.
\end{equation}
Derivations of the above two equations have been provided in many references dealing with quenches of harmonic oscillators, 
see e.g., \cite{spread1, Ali:2018fcz}. For details of these derivations, we refer the reader to these works and omit them here for brevity.

The operators $\mathbf{K}_{k}^{(+,-,0)}$ defined above 
 are related to the creation and annihilation operators before the  quench through the following relations
\begin{equation}\label{su11gene}
\begin{split}
&\mathbf{K}_k^+=\frac{1}{2}a_{0k}^{\dagger}a_{0k}^{\dagger}~,~~\mathbf{K}_k^0=\frac{1}{4} \Big(a_{0k}^{\dagger}a_{0k}+a_{0k}a_{0k}^{\dagger}\Big)~,\\
&\mathbf{K}_k^-=\frac{1}{2}a_{0k}a_{0k}~.
\end{split}
\end{equation}
Utilizing the standard commutation relations for the bosonic operators we can see that
operators  $\mathbf{K}_{k}^{(+,-,0)}$  provide a single-mode bosonic representation of the $su(1,1)$ Lie algebra,
thus they satisfy the following  commutation relations
\begin{equation}
\big[\mathbf{K}_{k}^+,\mathbf{K}_{k}^-\big]=-2\mathbf{K}_{k}^0~,~
~\big[\mathbf{K}_{k}^0,\mathbf{K}_{k}^{\pm}\big]=\pm \mathbf{K}_{k}^{\pm}~.
\end{equation}

The Casimir operator corresponding to the algebra, defined through the relation,
\begin{equation}
\mathbf{K}^2=\mathbf{K}_0^2-\frac{1}{2}\big(\mathbf{K}_+\mathbf{K}_-+\mathbf{K}_-\mathbf{K}_+\big)~,
\end{equation} 
commutes with all the three generators, and satisfies the eigenvalue equation
\begin{equation}\label{Casimir}
\mathbf{K}^2\big|j,h\rangle=h(h-1)\big|j,h\rangle~.
\end{equation}
The constant $h$ is known as the Bargmann index of the algebra, and $j$ takes values $0,1,2,\cdots$. For the single-mode bosonic 
representation of the $su(1,1)$ Lie algebra, $h$ can take values $1/4$ or $3/4$ (see \cite{Gerry:91}). In this paper we take $h$ to be $1/4$, 
for which the basis corresponding to a unitary irreducible representation of $su(1,1)$ algebra is the set of states with an even number of bosons.
Furthermore, the operations of the generators $\mathbf{K}_i$ on the states $\big|j,h\rangle$ are given by standard formulas, see, e.g. \cite{Gerry:91}.

From the identification that the operators $\mathbf{K}_{k}^{(+,-,0)}$ satisfy a $su(1,1)$ algebra, we see that the Hamiltonian
 after  the  quench  is actually an element of this algebra. Thus, the time-evolved state is a 
 generalised coherent state (CS) associated with
the $SU(1,1)$ group. Hence, the auto-correlation function (for each mode), given by
\begin{equation}\label{auto-cor}
\mathcal{S}_k(t)= ~_k\langle \Psi(t) \big| \Psi(t=0)\rangle_k = ~_k\langle 0\big|e^{i H_{1k} t}\big|0\rangle_k~,
\end{equation}
can be thought of as an overlap of a $SU(1,1)$ CS with the ground state 
before the quench \cite{Perelomov}. Our first goal in the rest of this section is to quantify the
spread of the time-evolved CS with respect to the state before quench in terms of Krylov basis elements.

We first  obtain  an analytical formula for the auto-correlation function defined above.  This is
 conveniently
done by using the standard decomposition formula for the $su(1,1)$ algebra (see e.g. \cite{Ban} for derivations 
of a collection of such well known relations).
Using  such decomposition formulas, we obtain the expression for time-evolved state to be 
\begin{equation}\label{evolved_state}
\begin{split}
\big| \Psi(t)\rangle_k =&
\exp\big[A_{1k}^+ (t) \mathbf{K}_k^+ \big]\exp\big[\ln \big(A_{1k}^0 (t)\big) \mathbf{K}_k^0 \big] \times\\
&\exp\big[A_{1k}^- (t) \mathbf{K}_k^- \big] \big|0\rangle_k
\end{split}
\end{equation}
The expressions for  time-dependent functions $A_{1}^0 (t)$ and $A_{1}^+ (t)$ appearing in the time-evolved state 
above are given by \footnote{\red{Since we are working with single mode wavefunctions, from now on we  remove the mode index $k$ for the rest of this section,
unless otherwise specified explicitly. The total complexity
is given by the  sum over all the modes.}}
\begin{equation}\label{coefficients}
\begin{split}
&A_{1}^0 (t) = \big[\cos (\omega_{1}t) + i \Omega_{1} \sin (\omega_{1} t)\big]^{-2} = f_1(t)^{-2}~,\\
&A_{1}^+ (t)=i \tilde{\Omega}_{1}\sin (\omega_{1} t) f_1(t)^{-1}~,
\end{split}
\end{equation}
where we have defined 
\begin{align}
\Omega_{1}=\frac{\omega_{0}^2 + \omega_{1}^2}{2 \omega_{0} \omega_{1}}~,~~\text{and}~~
\tilde{\Omega}_{1}=\frac{\omega_{0}^2 - \omega_{1}^2}{2 \omega_{0} \omega_{1}}~.
\end{align}
The time-evolved state  given in Eq. \eqref{evolved_state} can therefore  be written  in the form
\begin{equation}
\begin{split}
\big| \Psi(t)\rangle &= \big(A_1^0 (t)\big)^{1/4} \sum_{j=0}^{\infty} \frac{\big(A_1^+ (t)\big)^j}{j!}
( \mathbf{K}^+)^j \big|0\rangle~\\
&=\big(A_1^0 (t)\big)^{1/4} \sum_{j=0}^{\infty} 
\sqrt{\frac{\Gamma \big(j+\frac{1}{2}\big)}
{j! \sqrt{\pi}}} \big(A_1^+ (t)\big)^j  \big|j,h \rangle~,
\end{split}
\end{equation}
where in the last expression we have used the fact that here $h=1/4$, \red{and the states $ \big|j, h \rangle$
have been defined in Eq. \eqref{Casimir}}.
From this expression, it is then straightforward to obtain  the auto-correlation  function to be
\begin{equation}\label{ac1qu}
\mathcal{S}(t)=  \langle \Psi(t) \big| \Psi(t=0)\rangle  = \big(A_1^0 (t)\big)^{* 1/4} ~,
\end{equation}
with the expression for the time dependent function $A_1^0 (t)$ given in Eq. \eqref{coefficients}
above. Note that to obtain this 
auto-correlation function, we have neglected a phase factor which corresponds to the difference between the ground state 
energies before and after the quench. This factor is usually neglected in the discussion of the characteristics function
and the corresponding PDWD, as we have mentioned before.  Furthermore, using this auto-correlation function, or using the coherent 
state method of \cite{Caputa1, Balasubramanian:2022tpr}, we obtain the LC to be 
\begin{equation}
\begin{split}\label{LCchain}
a_n=\Big(2n+\frac{1}{2}\Big)\Omega_1~\omega_{1},~~n=0,1,2 \cdots\\
b_l= \frac{1}{2}\sqrt{2\big(2l^2-l\big)} \tilde{\Omega}_1 \omega_{1}~,~~l=1,2,3 \cdots~.
\end{split}
\end{equation}

\subsection{Evolution of the spread complexity}
Now since the post-quenched  time-evolved state is a $SU(1,1)$ CS,  to determine the expansion coefficients $\phi_n (t)$ required
in the computation of the SC,  we can use  the geometric method proposed in \cite{Caputa1}, \cite{Balasubramanian:2022tpr},
respectively in the context of the Krylov and spread complexity. Using the procedure explained in these references, we obtain the
exact expressions for the expansion coefficients to be 
\begin{equation}
\begin{split}
\phi_n (t) = &\mathcal{N}_{n} \phi_0 (t) \big(A_1^+ (t)\big)^n ~,~\text{with}~
\phi_0(t) = \big(A_1^0 (t)\big)^{1/4}~,\\
&\mathcal{N}_n= \sqrt{\frac{\Gamma \big(n+\frac{1}{2}\big)}
{n! \sqrt{\pi}}}~.
\end{split}
\end{equation}
The sum in the SC expression is performed 
exactly,  and the final expression for SC  of a single mode is given by \red{(here we have restored the mode index
$k$ to emphasize that this expression represents the SC of a single mode evolution)}
\begin{equation}\label{sc_single}
\mathcal{C}_k(t)=\frac{\big|\phi_{1k} (t)\big|^2}{\big(1- \mathcal{F}_k(t)\big)^{3/2}}~, ~~~\text{where}~~~~
\mathcal{F}_k(t) = \big|A^+_{1k} (t)\big|^2~.
\end{equation}
Using the expression for the $A^+_{1k} (t)$ given in Eq. \eqref{coefficients},
 we obtain the following simplified expression for the $k$th mode 
contribution to the total SC
\begin{equation}\label{sc1exact}
  \mathcal{C}_k(t)= \frac{1}{2}\tilde{\Omega}_{1k}^2 \sin ^2 (\omega_{1k} t)~
  =\frac{\big(\omega_{0k}^2 - \omega_{1k}^2\big)^2}{8 \omega_{0k}^2 \omega_{1k}^2}\sin ^2 (\omega_{1k} t)~.
\end{equation}

\subsection{Lanczos coefficients and the cumulants of the distribution of the work done}

We now illustrate the  general relation between the LC and the cumulants derived in section \ref{worklanczos}
for the single global quench of the harmonic chain considered in this section. In fact, since in this case the
analytical  expression for the 
auto-correlation function is relatively simple (see Eq. \eqref{ac1qu}), we can derive the first few relations
between the LC and the cumulants directly without using the general relations in  Eqs. \eqref{relation}
and \eqref{invrelation}. 

First consider the cumulant $\beta_1$, which from Eqs. \eqref{ac1qu} and \eqref{coefficients}, is
 obtained to be
equal to $\beta_1=\frac{1}{2}\omega_{1} \Omega_{1}$. Now from the expressions for $a_n$ given in Eq. \eqref{LCchain}
we see that this is exactly equal to the first of $a_n$s i.e., $\beta_1=a_0=\langle W\rangle$.\footnote{Note that here $\langle W\rangle$ 
is actually $\langle W_k\rangle$ i.e. work done on an individual mode. Since 
the harmonic chain is diagonalized in normal modes, total work is sum of works done on these individual modes.
This is true for higher moments as well. }
Similarly, we calculate the second cumulant from the expression for the auto-correlation function  
to be equal to $\beta_2=\frac{1}{2}\big(\Omega_{1}^2 -1\big)\omega_{1}^2=\frac{1}{2}\tilde{\Omega}_1^2 \omega_{1}^2$.
On the other hand, from the expression for the coefficients $b_n$, we see that
 $b_1^2=\beta_2=\langle W^2\rangle- \langle W\rangle^2$.
These two relations indicate an alternative interpretation for the LC in terms of the moments of the thermodynamic
quantity --  the work $W$ done on the system through the quench. In fact the first element of the first set of LC  $a_0$ 
is equal to the average work done on the system through the process of sudden quench, while the first element of the second set
of LC, i.e., $b_1$ is the standard deviation of this work done from the average value. Similarly, it is 
possible to write all the higher
LC as polynomials of the moments of the distribution of the work done with integer coefficients.
If we want to obtain the higher order relations, it is useful to directly use the general formula Eqs. \eqref{relation}, and \eqref{invrelation} given in the previous section.

$\bullet$ \textit{Interpretation for critical quenches.} 

To understand the above relations between the LC and various moments of the PDWD more clearly,
we consider a special case of the single quench scenario,
namely, we consider a critical quench where the final (or the initial) value of the frequency of the oscillators vanishes.
As discussed previously in \cite{Chandran:2022vrw,Pal:2022rqq, spread1}, when one considers a critical quench,
both the Nielsen complexity and the SC shows characteristic behaviours different from non-critical quenches. In particular, 
the divergence of the complexity at late times 
can be attributed  to the presence of the zero modes originated through the critical quench of the system \cite{Chandran:2022vrw}. 
Here we discuss its connection with the divergence of the average work done on the system
in such critical quenches.

First we write down the contribution of the $N$th mode towards the total SC. From Eq. \eqref{sc1exact}, we obtain this 
contribution to be (the subscript $N$ refers to the fact that these quantities correspond to mode number $N$)
\begin{equation}
\mathcal{C}_N(t)= \frac{1}{2}\tilde{\Omega}_{1N}^2 \sin ^2 (\omega_{1N} t)~
=\frac{\big(\omega_{0N}^2 - \omega_{1N}^2\big)^2}{8 \omega_{0N}^2 \omega_{1N}^2}\sin ^2 (\omega_{1N} t)~,
\end{equation}
where $\omega_{0N}$ and $\omega_{1N}$ are the $N$th mode normal frequencies before and after the quench.

We first consider the case when the frequency $\lambda_{1}$ of the individual oscillators vanishes after the
quench. In this case it is easy to see that the contribution of the $N$th mode (which is in fact a zero mode for
the critical quench) grows quadratically with time, i.e.,
\begin{equation}
\mathcal{C}_N(t)\big|_{\lambda_{1} \rightarrow 0}=\frac{1}{8}\lambda_{0}^2 t^2~.
\end{equation}
On the other hand, in the opposite case when the frequency before the  quench  is zero,  we see that the 
SC is divergent at all times. These behaviours of the SC for critical quenches  can be understood from the point of
view of the average work done on the system due to the quench. First we note that when the frequency before the 
quench vanishes, all the LC are divergent (see Eq. \eqref{LCchain}).
 Next, from the identifications $a_0=\langle W\rangle=\frac{1}{2}\omega_{1} \Omega_{1}$
and $b_1^2=\langle W^2\rangle- \langle W\rangle^2=\frac{1}{2}\tilde{\Omega}_1^2 \omega_{1}^2$, we see that in this case
the average work, as well as its variance diverge. This divergent behaviour of these two quantities are
very similar to what is observed when a quantum system  is quenched from criticality \cite{Paraan}.

The divergence of the average work done on the system (and the higher cumulants) is explained by observing that 
when there is a zero mode present in the system before the quench, it corresponds to a free particle, and the spread of the
initial wavefunction of this free particle in an $su(1,1)$ basis is infinity -- and hence the SC diverges.
Instead, when the system is quenched in such a way that the frequency after the quench is zero  it does not result in 
any divergence in the average work done on the system, since in this case the zero mode (i.e. the free particle)
results from an initially localised harmonic oscillator. 
Therefore, we conclude that when there is a zero mode present in the system before a sudden quench  it corresponds 
to a divergent average work as well as a divergent SC. We also note that the growth rate of the zero mode complexity
when $\lambda_1=0$ is proportional to the square of the initial frequency, whereas the average work corresponding to 
that mode is also proportional to the initial frequency. Hence, if we consider two different critical quench protocols 
where the initial frequencies are different, then the protocol with higher  frequency will correspond to greater average work
and higher growth rate of the SC. 
This discussion  thus provides a direct connection between the growth of SC in critical quenches and 
the average work done on the system.

\begin{figure}[h!]
	\centering
	\includegraphics[width=0.4\textwidth]{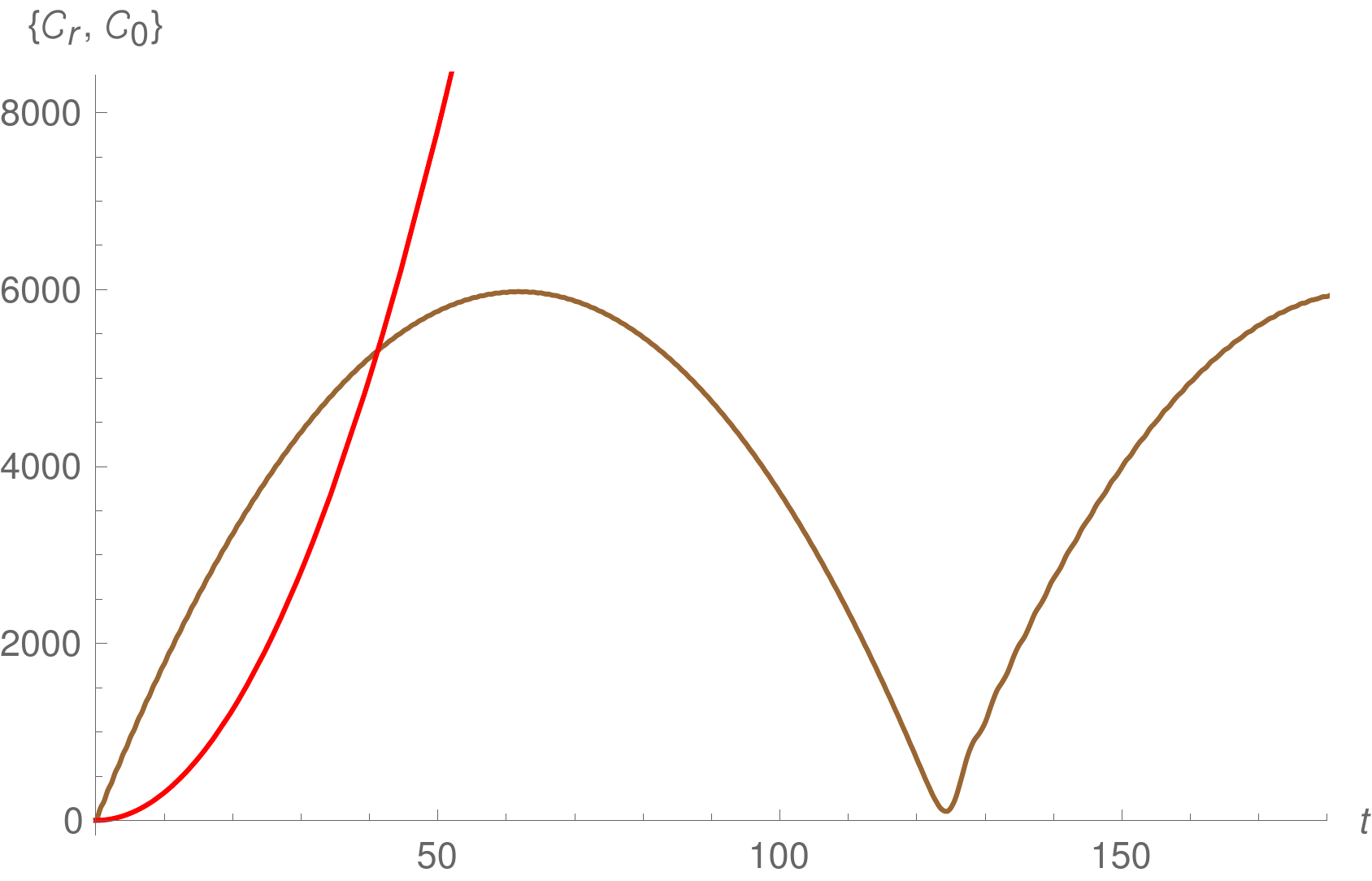}
	\caption{Time evolution of SC the zero mode (red) and the sum of the non-zero mode contributions (brown).
		Here $\lambda_0=5, k=2, N=350$. At large times the contribution of the zero mode always dominates.}
	\label{fig:SC_critical}
\end{figure}

Next, we graphically study the evolution of the SC for critical quench when the frequency after
the quench vanishes. In Fig. \ref{fig:SC_critical}, we have plotted the  contributions from  zero mode 
(denoted by $\mathcal{C}_0$) and sum of the rest of the modes (denoted by $\mathcal{C}_r$) towards the SC separately.
We see that, for early times, just after the quench, the contribution of the zero mode towards the total complexity 
is smaller than the some of the other modes. However, after a particular value of time (which depends on the initial
value of the frequency) the zero mode complexity becomes equal to the sum of other mode contributions and continues to
grow with time (whereas the the total complexity of the non-zero modes oscillates with time). The quadratic growth 
of the zero mode SC is therefore responsible for the overall quadratic growth of the SC with time.

Before concluding this section, we notice an important point regarding the nature of the time-evolved wavefunction,
the CS, and the associated LC. For the problem considered in this section -- quantum quench in a harmonic chain -- 
the time-evolved state 
is a generalised $SU(1,1)$ CS. Hence the associated Krylov basis is infinite dimensional, and we have an infinite 
number of LC $a_n$ and $b_n$. Here we have explicitly related only the first few LC with the cumulants of the 
work distribution.
On the other hand  when the time-evolved state belongs to a finite dimensional group, such as $SU(2)$, there 
are only finite numbers of Krylov basis elements and associated finite number of non-zero LC. 
In that case, it is possible to obtain explicit 
relations between all the LC with the cumulants of the work distribution.

\section{Mass quench of a bosonic scalar field in $d$-dimensions - Lanczos coefficients and  complexity}
\label{bosonic}

In this section, we consider the mass quench of a noninteracting bosonic model in $d$-spatial dimensions, and study
corresponding cumulants of the work distribution, the LC, and the relation between them in the limit that the linear 
size of the system $L \to \infty$. In this case, the general formula for the CF discussed in 
section \ref{ddimsys} is applied and the explicit time dependence of the function $f_C$ can be obtained.

The Hamiltonian of the system under consideration is that of a bosonic field $\phi$ of mass $m$ and is given by
the following expression when written in a diagonalized form in terms of  individual momentum modes \cite{Gambassi, Sotiriadis} 
\begin{equation}\label{scalar}
H (m)=\frac{1}{2}\int_k \big[\pi_k \pi_{-k}+\omega_k^2(m)\phi_k \phi_{-k}\big]~,
\end{equation}
where the modes $\phi_k$ and their conjugates satisfy the commutation relation 
$[\phi_k, \pi_{k^\prime}]=i (2\pi)^d \delta_{k, k^{\prime}}$, and $\omega_k(m)=\sqrt{k^2+m^2}$.
Furthermore, for the continuum model, the integral 
is over all the $d$ dimensional space $\int_k=\frac{\Omega_d}{(2\pi)^d} \int_{0}^{\infty} dk k^{d-1}$.
The above Hamiltonian  can be obtained, e.g., in the small interaction limit of the Sine-Gordon model --
one of the most popular models used to study nonlinearly interacting quantum systems.

We consider a mass quench in the model in Eq. \eqref{scalar}, where we change the mass $m_0$ of the field to a new value $m_1$
suddenly at $t=0$. This quench corresponds to a change in the dispersion relation, so that the  mode frequencies
before and after the quench are $\omega_{0k}$ and $\omega_{1k}$, respectively.
The CF corresponding to each mode (with momentum $k$) was obtained in \cite{Sotiriadis}, so that the CF 
is given by
\begin{equation}
\begin{split}
G(t)=&\prod_{k} G_{k}(t)=\prod_{k} e^{\frac{i}{2}(\omega_{0k}-\omega_{1k})}
\sqrt{\frac{1-\eta_k^2}{1-\eta_k^2 e^{-2i \omega_{1k}t}}}~,\\
&\text{where}~~\eta_k=\frac{\omega_{0k}-\omega_{1k}}{\omega_{0k}+\omega_{1k}}~.
\end{split}
\end{equation}
Notice that, apart from an unimportant phase factor, this CF for individual momentum modes is the same as the 
one we have derived  in Eq. \eqref{ac1qu} for a harmonic chain. The modulus squared
of the two expressions are  therefore identical.

Now taking the $L \to \infty$ limit of the logarithm of the CF, and replacing the sum over all the momentum
modes with an integral over $k$, we obtain
\begin{equation}
\begin{split}
\frac{\ln G(t)}{L^d}= &- \frac{it}{2} \int_k \big(\omega_{1k}-\omega_{0k}\big)+\frac{1}{2}\int_k (1-\eta_k^2) \\
&- \frac{1}{2} \int_k  \big(1-\eta_k^2 e^{-2i \omega_{1k} t}\big)~.
\end{split}
\end{equation}
Comparing this with the general expression for the CF for a general $d$-dimensional system of length $L$
given in Eq. \eqref{CF2}, we can easily identify the expressions for the  three functions $f_b$, $f_s$ and $f_{C}$.
As mentioned before, in the calculation of the cumulants and the moments, the contribution of the first term
coming from the phase factor in the CF will be neglected. 

The expressions for the first three cumulants 
calculated from the above CF (see Eq. \eqref{cumulants})  have been obtained in \cite{Sotiriadis} for 
different  spatial dimensions. For some values of $d$, these expressions are UV divergent. The average
of the work done on the system through the quench  is expressed as $\langle W\rangle=\langle 0\big|H(m)-H(m_0)\big|0\rangle$,
which apart from a trivial shift by an amount equal to the ground state energy of the pre-quench Hamiltonian, is just the 
first LC, i.e. $a_0$. Thus once again, like the case of harmonic chain considered in the previous section,
for the bosonic field quench set-up, the LC $a_0$ is the average work done on the system due to the 
quench. This is true for this model for any value of the  space dimension $d$.
 
Next, as it was noticed in \cite{Sotiriadis},
since $\ln G(t)$ is an extensive quantity  (it is proportional to $L^d$), all the cumulants calculated
from it using the  formula Eq. \eqref{CF2}, are also extensive. Thus we can define the probability distribution for
the intensive work $w=W/L^d$, and since all the higher order cumulants apart from $\beta_1$ and $\beta_2$
go to zero as $\mathcal{O}(\frac{1}{L^{2d}})$ or faster in the limit $L \rightarrow \infty $, the distribution $P(w)$ 
is  Gaussian, with mean $\beta_1=\gamma_1$ and variance $\beta_2=\gamma_2/L^d$, 
where $\gamma_1$ and $\gamma_2$ are constants, independent of $L$. The CF $\tilde{G}(t)$ corresponding to 
$P(w)$ is also Gaussian, and can thus be written as 
\begin{equation}\label{GaussianCF}
\tilde{G}(t)=\exp \bigg[-it \gamma_1 -\frac{t^2}{2 L^d} \gamma_2 +\mathcal{O}\Big(\frac{1}{L^{2d}}\Big)\bigg]~.
\end{equation}

Now using this form for the CF and the explicit expressions for the complete Bell polynomials \cite{Comtet} in Eq. \eqref{invrelation},
we can obtain the moments of the distribution of the intensive work done on the system. To obtain a compact expression for 
these moments in the $L \rightarrow \infty$ limit, it is instructive first consider expressions for the Bell polynomials
$Y_n$ as a function of $\beta_1, \beta_2, \cdots \beta_n$. 
For example, we have $Y_4(\beta_1, \beta_2,\beta_3, \beta_4)=\beta_1^4+6\beta_1^2 \beta_2 + 4\beta_1 \beta_3+
3 \beta_2^2+\beta_4$, and $Y_5(\beta_1, \beta_2,\beta_3, \beta_4, \beta_5)=\beta_1^5+10\beta_2 \beta_1^3+15\beta_2^2\beta_1
+10\beta_3^3\beta_1^2+10\beta_3\beta_2+5\beta_4\beta_1+\beta_5$. From these two expressions,  
as well as the expressions  for the higher order Bell polynomials, we see that there are  only two terms which are either
$\mathcal{O}(1)$ or $\mathcal{O}(1/L^d)$, i.e. survive in the limit $L \rightarrow \infty$. The first one is
equal to $\beta_1^n$, and the second one is proportional to $\beta_2$, and has coefficient $\mathcal{N}_{n-2} \beta_1^{n-2}$, 
for a sequence of positive numbers $\mathcal{N}_n$. Apart from these two terms,
all the other terms in the expression for the complete Bell polynomials $Y_n$  are $\mathcal{O}(\frac{1}{L^{2d}})$ or smaller, 
and hence, it is possible to  neglect them in the $L \rightarrow \infty$ limit.

Therefore, from Eq. \eqref{invrelation} we have expression for the moments to be 
\begin{equation}
M_n= (-i)^n \Big[\gamma_1^n+\frac{\mathcal{N}_{n-2}}{L^d} \gamma_1^{n-2}\gamma_2\Big]~,
\end{equation}
where, the expression for the polynomial $\mathcal{N}_n$ is given by $\mathcal{N}_n=\frac{1}{2}\big(n^2+2n+3\big)$.

Next, to find out the LC corresponding to these moments and the auto-correlation function $\tilde{G}(t)^*$ 
(with $\tilde{G}(t)$ given in Eq. \eqref{GaussianCF} above), we note that,
as observed in \cite{Caputa1} and \cite{Balasubramanian:2022tpr},  the Gaussian CF corresponds to a particle moving in 
the Weyl-Heisenberg group with the Hamiltonian of the form 
\begin{equation}
H=\Lambda (a^\dagger +a) +\Omega a^\dagger a+\delta \mathbf{I}~,
\end{equation}
where $\Lambda$, $\Omega$ and  $\delta$ are constants. When $\Omega \ll 1$, so that in the CF contributions 
of $\mathcal{O}(\Omega^2)$ and higher can be neglected, then the CF 
(of the particle moving in the Weyl-Heisenberg group), which is equal to the coefficient of $\big|K_0\rangle$ 
in the expansion of the time-evolved state in terms of the Krylov basis, is a Gaussian function of time of the form given
in Eq. \eqref{GaussianCF}. In our case, we can identify, $\delta=\gamma_1$ and $\Lambda=\sqrt{\frac{\gamma_2}{L^d}}$.
Furthermore, here $\Omega \approx \mathcal{O}\big(\frac{1}{L^{2d}}\big)$, and 
hence in the limit $L \rightarrow \infty$, $\Omega \rightarrow 0$.
In this limit, it is possible to write the 
 LC approximately as  \cite{Caputa1, Balasubramanian:2022tpr}
\begin{equation}
a_n\approx \gamma_1~, ~~~\text{and}~~~b_n=\sqrt{\frac{\gamma_2 n}{L^d}}~.
\end{equation}
In the exact expressions for the LC $a_n$s for a particle moving in the Weyl-Heisenberg group, 
there is an additional additive term, which increases with $n$. 
However, since this term is proportional to $\Omega$, we have neglected it here, and the $a_n$ coefficients are just constants.
This term will contribute only at the $n \rightarrow \infty$ limit. 

Now the SC of this bosonic field system after the mass quench in the limit $L \rightarrow \infty$ can also be obtained
by taking the $\Omega \rightarrow 0$ limit of the general formula for the SC for a particle moving in the Weyl-Heisenberg group.
We obtain the corresponding expression for the SC to be
\begin{equation}
\mathcal{C}(t)\Big|_{L \rightarrow \infty } \approx \Lambda ^2 t^2 = \frac{\gamma_2}{L^d} t^2 ~.
\end{equation}
This is the leading contribution to the SC, with all the subleading contributions being $\mathcal{O}\big(\frac{1}{L^{2d}}\big) $ 
and smaller, and thus SC  grows quadratically with time after quench in non-interacting bosonic field 
theory. 
However, since $\gamma_2$ is finite and we are considering the limit $L \rightarrow \infty$,
this quadratic growth will be apparent only when we study the system a long time after the quench. 
Furthermore, for higher spatial 
dimension of the  system, this growth may be smaller than that with lower spatial dimension, even at late times. This example 
therefore makes it clear that whenever the 
PDWD of a system due to a sudden quench is Gaussian, the associated CF is also a Gaussian function of time, so that
the corresponding SC grows quadratically with time, 
with the coefficient of the growth being proportional to the variance of the distribution.
This conclusion  is true irrespective of the details of the system under consideration.

This example also nicely illustrates the usefulness of connecting the  SC with the PDWD.  To reiterate,
	due to the close relationship between the  CF and 
	work distribution, the former is fixed whenever the  distribution of  work under a quench 
	is specified.  Hence, 
	the LC and the SC obtained from the  CF are also determined, and can only have the same form
	for different systems that might have similar work distribution under a quench. Furthermore,  this  simple 
	example of quench in a non-interacting bosonic field theory  can be used as the starting
	point for studying the SC evolution in quenches of general interacting field theories. Presumably, 
	in presence of interaction, one  needs to consider 
	work distribution which deviates from the Gaussian one for the non-interacting case, and SC will 
	grow with a more intricate pattern compared to the simple quadratic growth obtained here. This is an interesting
	problem, and we leave it for a future study.

\section{Conclusions and discussions}
\label{conclusions}

In this paper, we have provided a generic relation between the statistics of work done on a quantum system under a sudden 
quantum quench and the LC associated with the Krylov basis constructed using the post-quench Hamiltonian.
By using the relation between moments of the auto-correlation function and the corresponding cumulants of the 
probability distribution, we have shown that it is possible to express the LC  
in terms of the  physically measurable 
quantities, such as the average, variance, and higher order cumulants of the  work done on the system through the quench. 
We believe that this should be an important step towards understanding the significance of these coefficients, specifically in a quench scenario,
and  circuit evolution in general.

We have applied our findings to two realistic examples, the first being the time evolution of the SC in a quenched harmonic chain
with nearest neighbour interaction. We have shown that $a_{0}$ is equal to  average of the work done on the system, 
while $b_{1}$ represents the standard deviation of this work from the corresponding average. 
Using this observation, we can explain the  fact that SC for time evolution under a critically 
quenched Hamiltonian diverges at all times, since the corresponding average work diverges as well.

Similarly, for the second example we consider -- a mass quench in a bosonic scalar field theory 
in the limit of large system size, 
we verified the same relation between  the LC and cumulants of the work done on the system. 
Since in this limit, the probability distribution is Gaussian,  the SC is seen to be growing 
quadratically with time. As we have discussed, this feature shown by the SC is true whenever this distribution,
and hence the CF is Gaussian.

 We conclude by pointing out a few potential future applications of the results presented here. Firstly, as discussed in the introduction,
	our goal is to provide a unified description of observables studied in quantum quenches by relating quantities from two different sets of such observables. Here we have 
	illustrated one such connection 
	between a  thermodynamic quantity  (work distribution) and an information theoretic measure (complexity of spread of a time-evolved state) by relating 
	the moments of the work distribution with the LC 
	corresponding to the post-quench Hamiltonian evolution. As an application, these relations can be useful to understand the 
	characteristics features of  SC evolution and the pattern of the LC 
	in a chaotic system \cite{Balasubramanian:2022tpr}, as well as the phenomena of information scrambling
	in such systems. Furthermore, it will be interesting to see whether similar relations can be established
  between other information theoretic quantities which are commonly studied in quenches, 
  such as the entanglement entropy or the out-of-time order 
  correlator, and thermodynamic quantities, e.g.,  the entropy or the heat generated, etc. 

Secondly, in this paper we have considered sudden quenches
	in bosonic systems which can be diagonalised via normal modes and, therefore, are non-interacting in nature. As a future application, one can consider quenches in realistic 
	interacting many-body quantum systems, and quenches in fermionic field theories, and see whether the kind of relations between the work statistics and the LC that we have established here are also valid in more general quench scenarios as well \cite{Gautam:2023bcm}. This should further our understanding of the SC and LC in terms of thermodynamic quantities.

Finally, the results presented  in this paper offer the first steps of understanding the significance of the LC, as well as the 
	SC, for time evolution after a sudden quantum quench. Though here we have considered time evolution after a quantum 
	quench only,  a similar construction can be envisaged for any general circuit evolution as well. In  this case, the 
	relation between LC and the quantities analogous to the average and  variance of the work done can shed 
	light on understanding the link between the Krylov basis construction, the SC, and the geometric formulation of 
	circuit complexity \cite{Nielsen1,Nielsen2, Nielsen3,Jefferson:2017sdb}. We hope to report on this in the near future \cite{Work}. 

\begin{center}
	\bf{Acknowledgements}
\end{center}
We thank our anonymous referees for their constructive
comments and criticisms which helped to improve a draft
version of this manuscript. The work of TS is supported in part by the USV Chair
Professor position at the Indian Institute of Technology,
Kanpur.

\appendix

\section{Moments for quench in a single harmonic oscillator}\label{homoment}
In this appendix we illustrate the use of the general formula for the moments in Eq. \eqref{genmoments} 
for quench in a single harmonic
oscillator with frequency $\omega_{0}$ to a new frequency $\omega$. This can then be easily generalised to the case
of the quench in a harmonic chain discussed in section \ref{singlequench}.

To use the formula in Eq. \eqref{genmoments} we need to evaluate the overlap between the ground state before the quench
and an arbitrary  number state $\big|\tilde{n}\big>$  after the quench. For a quench in a harmonic oscillator, this 
is given by  \cite{Sotiriadis}
\begin{equation}\label{hooverlap}
\begin{split}
\big<\tilde{n}\big|0\big> = \left\{\begin{array}{ll} \Big[\frac{(-\eta)^n}{\mathcal{U}} \frac{(n-1)!!}{n!!}\Big]^{1/2}~~~~\text{for}~~~n=0,2,4 \cdots\\
0  \quad \qquad \qquad~~ \text{for}~~~~n =1,3,5 \cdots.~\end{array}\right.
\end{split}
\end{equation}
Here we have defined,\footnote{For convenience here we assume that $\omega_{0}<\omega$, i.e., $\eta$ is negative.}  
\begin{equation}
\eta=\frac{\omega_{0}-\omega}{\omega_0+\omega}~,~~~\text{and}~~~ \mathcal{U}=\frac{\omega_0+\omega}{2\sqrt{\omega_{0} \omega}}~.
\end{equation}
Substituting this overlap in Eq. \eqref{genmoments}, we have the expression for the moments
of the work distribution for a quench in a single harmonic oscillator as
\begin{equation}\label{momentsho}
M_n=(-i)^n \sum_{j=even}  \bigg[\frac{(\eta)^j}{\mathcal{U}} \frac{(j-1)!!}{j!!}\bigg] \Big(\big(j+\frac{1}{2}\big)\omega 
-\frac{1}{2}\omega_{0}\Big)^n~.
\end{equation}

\section{Derivation  of the expression for $a_1$}\label{a_1der}
In this Appendix we briefly describe the derivation of the expressions in  Eqs. \eqref{a0b1} and
\eqref{a_1} for the first three LC
in terms of averages of various powers of  the work done $W$. First, we write down the expressions for $a_0$,
$a_1$ and $b_1$
in terms of the moments of the work distribution
\begin{equation}\label{a_1m}
a_0=iM_1~, ~~ b_1^2=-M_2+M_1^2~,~~a_1=\frac{iM_1^3-iM_3}{M_1^2-M_2}-2iM_1~.
\end{equation}
We have taken these  expressions from well known tabulated relations between 
the moments of the LC (see, e.g., 
the Table 4.2 on page - 37 of \cite{Viswanath}).
Now, from Eq. \eqref{CF}, since $G(t)=\big<e^{-i W t}\big>$, we get the expressions for the first 
three moments to be  
\begin{equation}
M_1=-i\big<W\big>~,~~~ M_2=-\big<W^2\big>~, ~~\text{and}~~M_3=i\big<W^3\big>~.
\end{equation}
Substituting these in Eq. \eqref{a_1m} above, we get the relations given in Eqs. \eqref{a0b1} and
\eqref{a_1}. An entirely similar procedure can be used to obtain the expression for 
other LC in terms of  averages of various powers of the work done using the tabulated relations 
between the LC and the moments of the CF.


\end{document}